\documentclass[conference]{IEEEtran}
\IEEEoverridecommandlockouts

\usepackage{cite}
\usepackage{amsmath,amssymb}
\usepackage{graphicx}
\usepackage{booktabs}
\usepackage{url}
\usepackage{xurl}
\usepackage[hidelinks]{hyperref}

\begin{document}

\title{Pneumonia Detection on Chest X-Rays Derived from a U-Net Segmentation Model}

\author{\IEEEauthorblockN{Manasvini Singh}
\IEEEauthorblockA{Campbell, CA, USA\\
manasvinisingh520@gmail.com}}

\maketitle

\begin{abstract}
Pneumonia killed more than 740{,}000 children under the age of five in 2019,
and it can be difficult to recognize on chest radiographs (CXRs): opacities
may be faint in early disease, may sit behind the equally bright heart
shadow, and may resemble other lung conditions. Most deep learning systems
treat pneumonia detection as a yes-or-no classification task. This paper
asks whether a model trained on a harder task, segmenting pneumonia pixel by
pixel, can be turned into an effective detector for the simpler task of
deciding whether an image contains pneumonia at all. The hypothesis is that
the detailed supervision of segmentation forces the network to learn where
and what the disease looks like, producing a better detector than training
for detection alone. A U-Net was trained to segment pneumonia on the RSNA
Pneumonia Detection Challenge dataset of 26{,}684 radiographs, whose
rectangular bounding boxes were converted into pixel masks. The data are
heavily imbalanced: 22.5\% of images are positive, and only 2.6\% of pixels
fall inside a box, so focal loss with $\alpha = 0.25$ and $\gamma = 2$ and
dropout were needed for the model to learn. On the harder segmentation task,
the model reached a pixel-level F1 score of 0.45 on a held-out test set of
2{,}668 images. Applied to the simpler detection task, the same model
reached a precision of 0.71, recall of 0.52, and F1 score of 0.60, with 91\%
specificity, and consistent scores on the development set. These results
show that a segmentation-trained model transfers to detection while also
showing where the suspected disease is. A direct comparison with a
classifier trained only for detection on the same data is needed to confirm
whether the segmentation-first approach is better, and is the main
direction for future work.
\end{abstract}

\section{Introduction}

\subsection{Motivation}

Pneumonia is an infection of the lungs that occurs in every age group but
is especially dangerous for young children. According to the World Health
Organization, it killed 740{,}180 children under the age of five in 2019,
accounting for 14\% of all deaths in that age group \cite{who}. Prompt
diagnosis matters because effective treatment is widely available.

Chest radiographs (CXRs) are one of the main tools radiologists use to
identify pneumonia. A CXR is a grayscale image in which dense structures,
such as bone, the heart, and fluid, appear bright, while air-filled lung
appears dark. Pneumonia typically appears as an area of increased opacity:
a region of the lung that looks whiter and less transparent than it should.
Reading these images is difficult for several reasons. Early disease may
produce only faint changes. An opacity located behind the heart overlaps a
region that is already bright, so the radiologist is effectively looking
for white on white. Pneumonia can also resemble other conditions, such as
bronchitis, and radiologists do not always agree on whether a radiograph
shows pneumonia \cite{neuman2012}.

\subsection{Research Question}

Many deep learning systems treat pneumonia detection as image
classification: the network sees a radiograph and is trained only to output
a single yes-or-no answer or probability. Such a model receives very little
information from each training example, one bit per image, and nothing in
that training signal tells it where the disease is.

This project tests a different strategy: train the model on a harder task
first, then use it for the simpler one. A segmentation network is trained
to predict, for every pixel, whether it belongs to a region of pneumonia.
This task is more demanding, because the model must find the disease and
mark its extent, but it also provides far richer supervision: every pixel
of every training image carries a label. The trained segmentation model is
then used for detection by calling an image positive when its predicted
mask contains pneumonia. The question this paper investigates is:

\begin{quote}
\emph{Does a model trained to segment pneumonia make a good pneumonia
detector, and can learning the harder task produce better detection than
training for detection alone?}
\end{quote}

The hypothesis is that it can. To segment pneumonia well, a model has to
learn what opacities look like and where they occur in the lungs, so a
model that has mastered the harder task should find the simpler one
comparatively easy. A segmentation-based detector has a second advantage:
alongside its diagnosis it shows where it believes the disease is, which a
radiologist can check.

\subsection{Approach}

The segmentation model is a U-Net \cite{ronneberger2015}, an architecture
designed for biomedical image segmentation. It was trained on the RSNA
Pneumonia Detection Challenge dataset \cite{shih2019,rsnachallenge}, whose
positive images are annotated with rectangular bounding boxes rather than
precise outlines. The boxes were treated as approximate pixel masks, which
limits how precise the segmentation can be but makes it possible to train a
segmentation model on a large, expert-labeled dataset. The model was then
evaluated on both tasks: segmentation, scored pixel by pixel, and
detection, scored image by image.

\subsection{Contributions}

\begin{itemize}
\item A segmentation-first approach to pneumonia detection, in which a
U-Net trained on bounding-box masks is used to diagnose whole images.
\item An evaluation of the same model on both the harder segmentation task
and the simpler detection task, including the full image-level confusion
matrix and an analysis of its errors.
\item A comparison of loss functions and dropout settings under severe
pixel-level class imbalance, showing that binary cross-entropy fails and
focal loss allows the minority class to be learned.
\end{itemize}

\section{Related Work}

\subsection{Pneumonia Classification on CXRs}

Large public datasets have made deep learning on CXRs practical. Wang
\emph{et al.}~\cite{wang2017} released ChestX-ray8, a set of more than
100{,}000 frontal radiographs from the U.S. National Institutes of Health
(NIH) with disease labels mined from radiology reports. Building on its
expanded version, Rajpurkar \emph{et al.}~\cite{rajpurkar2017} trained
CheXNet, a 121-layer convolutional network, and compared it with four
practicing radiologists on 420 test images. CheXNet reached an F1 score of
0.435 for pneumonia, against a radiologist average of 0.387. The low scores
for both reflect how ambiguous pneumonia is on radiographs. Kermany
\emph{et al.}~\cite{kermany2018} applied transfer learning to classify
pediatric CXRs as normal or pneumonia.

\subsection{The RSNA Pneumonia Detection Challenge}

The Radiological Society of North America (RSNA), the Society of Thoracic
Radiology, and MD.ai re-annotated a subset of the NIH images for possible
pneumonia \cite{shih2019}. Radiologists drew bounding boxes around
opacities suspicious for pneumonia, and the dataset was released as the
2018 RSNA Pneumonia Detection Challenge, in which more than 1{,}400 teams
took part \cite{rsnachallenge}. Entries were scored by mean average
precision over intersection-over-union (IoU) thresholds from 0.4 to 0.75.
The best private-leaderboard scores were about 0.25, and top entries used
object detectors such as RetinaNet with large pretrained backbones and
model ensembles \cite{gabruseva2020}. The low absolute scores again
indicate how hard it is to localize pneumonia consistently.

\subsection{Segmentation and Class Imbalance}

U-Net \cite{ronneberger2015} combines a contracting path, which captures
context, with an expanding path that recovers spatial detail, and connects
the two with skip connections. It was designed to work well with limited
training data and has become a standard architecture for medical image
segmentation. In segmentation tasks where the target occupies a small
fraction of the image, standard cross-entropy loss is dominated by the
many easy background pixels. Focal loss \cite{lin2017} addresses this by
down-weighting well-classified examples so that training concentrates on
the hard, rare ones.

\section{Dataset}

\subsection{RSNA Pneumonia Detection Dataset}

This study used the training images of the RSNA Pneumonia Detection
Challenge \cite{rsnachallenge}: 26{,}684 frontal CXRs, each from a
different patient, of which 15{,}166 are from male patients and 11{,}518
from female patients. Of these, 6{,}012 images (22.5\%) contain at least
one bounding box marking a lung opacity suspicious for pneumonia; the
remaining 20{,}672 are negative. Among positive patients, the most common
age range for both sexes was 50 to 60 years.

\begin{figure}[t]
\centering
\includegraphics[width=\columnwidth]{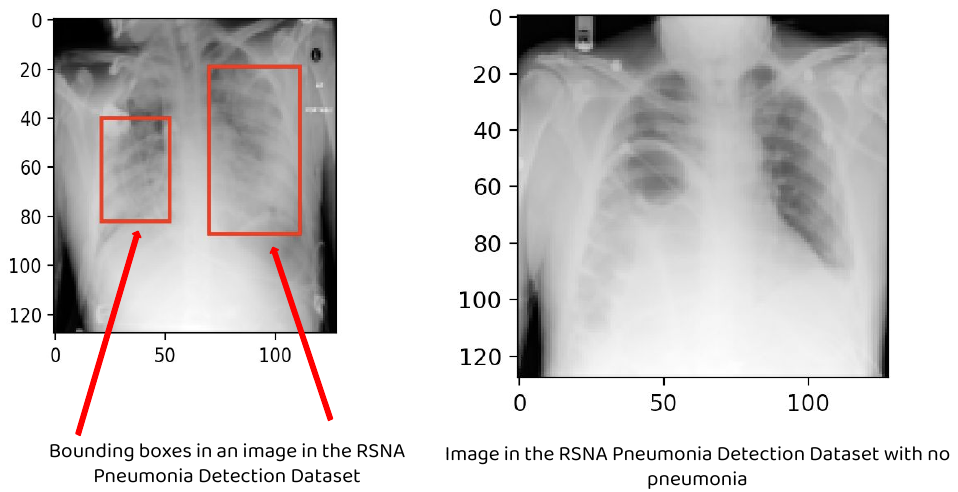}
\caption{Examples from the RSNA dataset after resizing to $128\times128$
pixels. Left: a positive image with two radiologist-drawn bounding boxes.
Right: an image with no pneumonia.}
\label{fig:rsna}
\end{figure}

Fig.~\ref{fig:rsna} shows a positive image with its bounding boxes and a
negative image.

\subsection{Preprocessing}

Images were resized to $128\times128$ pixels. For each image, a binary
target mask of the same size was created by setting every pixel inside a
bounding box to 1 (pneumonia) and all other pixels to 0. Negative images
received an all-zero mask.


\subsection{Class Imbalance}

The resulting data are imbalanced at two levels (Table~\ref{tab:imbalance}).
At the image level, fewer than a quarter of images are positive. At the
pixel level, the imbalance is far more severe: even in positive images,
bounding boxes cover only part of the lungs, and across the whole dataset
only 2.6\% of pixels are labeled as pneumonia. A model that labels every
pixel as background would therefore achieve over 97\% pixel accuracy while
detecting nothing, so accuracy is not a useful measure for this task.

\begin{table}[t]
\centering
\caption{Class Balance in the RSNA Dataset at $128\times128$ Pixels}
\label{tab:imbalance}
\begin{tabular}{lrrr}
\toprule
Level & Positive & Total & Positive share \\
\midrule
Images & 6{,}012 & 26{,}684 & 22.5\% \\
Pixels & 11{,}290{,}760 & 437{,}190{,}656 & 2.6\% \\
\bottomrule
\end{tabular}
\end{table}

\subsection{Splits}

The images were divided into training, development (validation), and test
sets. The training set was used to fit the models, the development set to
compare configurations, and the test set only for final evaluation. The
test set contains 2{,}668 images (10\% of the data), of which 823 are
positive.


\section{Method}

\subsection{Network Architecture}

\begin{figure*}[t]
\centering
\includegraphics[width=0.9\textwidth]{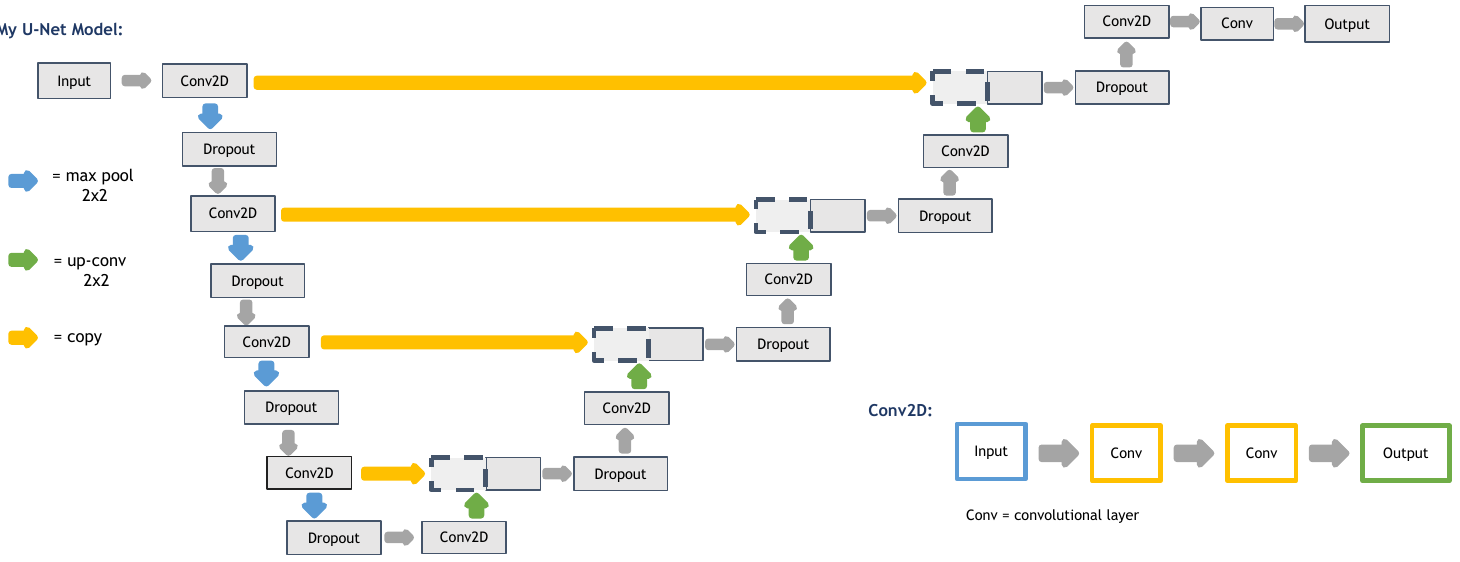}
\caption{U-Net architecture used in this work. Each Conv2D block consists
of two convolutional layers. Blue arrows denote $2\times2$ max pooling,
green arrows $2\times2$ up-convolution, and yellow arrows skip connections
that copy feature maps from the contracting path to the expanding path.}
\label{fig:unet}
\end{figure*}

The network follows the U-Net design \cite{ronneberger2015}
(Fig.~\ref{fig:unet}). Its basic unit, labeled Conv2D in the figure, is a
block of two consecutive convolutional layers. The contracting path applies
four such blocks, each followed by dropout and $2\times2$ max pooling, which
halves the spatial resolution. A fifth block forms the bottleneck. The
expanding path mirrors the contracting path: at each of four levels, a
$2\times2$ up-convolution doubles the resolution, the result is concatenated
with the feature map from the same level of the contracting path through a
skip connection, and dropout and a convolutional block follow. A final
convolutional layer produces a single-channel map giving, for each pixel,
the predicted probability of pneumonia. In the final model, the first block
has 16 filters.


\subsection{Loss Function}

The first model was trained with binary cross-entropy. Because almost all
pixels are background, this loss can be made small by predicting
background everywhere, and the model did not learn to detect pneumonia
(Section~\ref{sec:experiments}). Subsequent models used focal loss
\cite{lin2017},
\begin{equation}
\mathrm{FL}(p_t) = -\alpha_t \, (1 - p_t)^{\gamma} \log(p_t),
\end{equation}
where $p_t$ is the predicted probability of the true class for a pixel,
$\alpha_t$ is a class-weighting factor, and $\gamma \ge 0$ controls how
strongly well-classified pixels are down-weighted. With $\gamma = 0$, focal
loss reduces to weighted cross-entropy. As $\gamma$ grows, pixels the model
already classifies confidently contribute less to the loss, so training
focuses on difficult pixels, which are disproportionately the rare
pneumonia pixels and their boundaries.

\subsection{From Segmentation to Detection}

To diagnose an image, the predicted probability map was thresholded to
obtain a binary mask, and the image was classified as positive when the
mask contained pneumonia pixels and negative otherwise.


\subsection{Evaluation Metrics}

Because of the class imbalance, performance is reported with precision,
recall, and the F1 score:
\begin{equation}
P = \frac{TP}{TP + FP}, \quad
R = \frac{TP}{TP + FN}, \quad
F_1 = \frac{2PR}{P + R},
\end{equation}
where $TP$, $FP$, and $FN$ are the numbers of true positives, false
positives, and false negatives. These metrics were computed at two levels.
At the \emph{pixel level} (segmentation), each pixel is one sample. At the
\emph{image level} (detection), each radiograph is one sample. During
model development, the IoU between predicted and target masks was also
monitored.

\section{Experiments}
\label{sec:experiments}

\subsection{Choice of Loss Function}

A model trained with binary cross-entropy (\texttt{binary\_accuracy\_m1}
in Table~\ref{tab:explore}) did not produce a usable F1 score on either the
training or development set, consistent with collapsing to the majority
class. Switching to focal loss (\texttt{focal\_loss\_m2}) produced a model
that fit the training set well (F1 = 0.87) but generalized poorly
(development F1 = 0.27).

\subsection{Focal-Loss Parameters}

Several combinations of $\alpha$ and $\gamma$, chosen at random, were
compared by how quickly they reduced the loss and by the training IoU they
reached. Several combinations caused training to collapse to an IoU of
zero. The setting $\alpha = 0.25$, $\gamma = 2.0$, which is also the
default recommended by the original focal-loss paper \cite{lin2017},
reached a training IoU of 0.87 and was used in all subsequent models.

\subsection{Dropout and Model Capacity}

All early focal-loss models overfit: they scored far higher on the
training set than on the development set. Dropout was therefore added and
varied. Table~\ref{tab:dropout} shows a sweep over dropout rates. Without
dropout, the model nearly memorized the training masks (IoU 0.99) but
reached only 0.19 IoU on the development set. A dropout rate of 0.8
prevented the model from learning at all. Applying dropout of 0.2 inside
the convolutional blocks gave the best development scores. Table~\ref{tab:explore}
summarizes the other configurations explored, including a version with 32
initial filters.

\begin{table}[t]
\centering
\caption{Dropout Sweep (Focal Loss)}
\label{tab:dropout}
\small\setlength{\tabcolsep}{4pt}
\begin{tabular}{lcccc}
\toprule
Model & Train IoU & Dev IoU & Dev F1 & Epochs \\
\midrule
\texttt{dropout\_p0} & 0.99 & 0.19 & 0.33 & 50 \\
\texttt{dropout\_p1} & 0.97 & 0.20 & 0.33 & 50 \\
\texttt{dropout\_p3} & 0.90 & 0.21 & 0.35 & 50 \\
\texttt{dropout\_p8} & 0.45 & 0.05 & 0.10 & 50 \\
\texttt{conv2d\_drop\_p2} & 0.82 & 0.23 & 0.38 & 20 \\
\bottomrule
\multicolumn{5}{p{0.9\columnwidth}}{\footnotesize The suffix gives the
dropout rate (p3 = 0.3). In \texttt{conv2d\_drop\_p2}, dropout of 0.2 is
applied within the convolutional blocks.}
\end{tabular}
\end{table}


\begin{table}[t]
\centering
\caption{Pixel-Level F1 of Explored Configurations}
\label{tab:explore}
\small\setlength{\tabcolsep}{4pt}
\begin{tabular}{lcc}
\toprule
Model & Train F1 & Dev F1 \\
\midrule
\texttt{binary\_accuracy\_m1} (cross-entropy) & -- & -- \\
\texttt{focal\_loss\_m2} & 0.87 & 0.27 \\
\texttt{alpha\_gamma\_m3} ($\alpha{=}0.25$, $\gamma{=}2$) & 0.88 & 0.27 \\
\texttt{change\_dropout\_m5} (dropout 0.2) & 0.61 & 0.40 \\
\texttt{new\_dropout\_p4} & 0.27 & 0.23 \\
\texttt{new\_no\_filters\_32} (dropout 0.4) & 0.68 & 0.39 \\
\bottomrule
\multicolumn{3}{p{0.9\columnwidth}}{\footnotesize -- : training did not
produce a valid F1 score.}
\end{tabular}
\end{table}

\subsection{Final Model Selection}

The three strongest configurations were retrained and compared
(Table~\ref{tab:top3}). Model~3 scored highest on the training set but
showed the largest gap between training and development performance.
Model~1 had the highest development F1 (0.40) and the smallest gap, and
was selected as the final model. Its settings are a dropout rate of 0.2,
batch size of 8, focal loss with $\alpha = 0.25$ and $\gamma = 2.0$, a
learning rate of 0.001, and 16 initial filters.

\begin{table}[t]
\centering
\caption{Top Three Configurations (Pixel-Level F1)}
\label{tab:top3}
\small\setlength{\tabcolsep}{4pt}
\begin{tabular}{p{0.58\columnwidth}cc}
\toprule
Configuration & Train & Dev \\
\midrule
Model 1: dropout 0.2, batch 8, lr 0.001, 16 filters & 0.61 & \textbf{0.40} \\
Model 2: dropout 0.2, batch 8, lr 0.001, 32 filters & 0.68 & 0.39 \\
Model 3: dropout 0.2 in conv blocks only, batch 16, lr 0.0001, 16 filters & 0.77 & 0.38 \\
\bottomrule
\multicolumn{3}{p{0.9\columnwidth}}{\footnotesize All use focal loss with
$\alpha = 0.25$, $\gamma = 2.0$.}
\end{tabular}
\end{table}


\section{Results}

\subsection{Performance on the Harder and Simpler Tasks}

Table~\ref{tab:results} reports the final model's performance on both tasks.
On the harder segmentation task, the model reached pixel-level F1 scores
of 0.40 and 0.45 on the development and test sets. On the test set, precision was 0.71 but recall only
0.33: when the model marked a pixel as pneumonia it was usually inside a
box, but it marked only a third of the boxed pixels.

When the same model was applied to the simpler detection task, it reached
an F1 score of 0.60 on both the development and test sets. The detection
scores were identical across the two sets, which suggests that the
segmentation model's ability to recognize pneumonia generalizes stably to
unseen images.

\begin{table}[t]
\centering
\caption{Final Model Performance}
\label{tab:results}
\begin{tabular}{llccc}
\toprule
Task & Set & Precision & Recall & F1 \\
\midrule
Segmentation (pixels) & Dev  & 0.51 & 0.33 & 0.40 \\
                      & Test & 0.71 & 0.33 & 0.45 \\
\midrule
Detection (images)    & Dev  & 0.71 & 0.52 & 0.60 \\
                      & Test & 0.71 & 0.52 & 0.60 \\
\bottomrule
\end{tabular}
\end{table}

\subsection{Image-Level Errors}

Table~\ref{tab:confusion} gives the test-set confusion matrix for
detection. Of the 1{,}845 negative images, 1{,}672 were correctly
identified, a specificity of 0.91. Of the 823 positive images, 431 were
detected, a sensitivity (recall) of 0.52. Overall accuracy was 0.79, which
is only modestly above the 0.69 that would be obtained by labeling every
test image negative. This comparison illustrates why F1 rather than
accuracy was used to select models. When the model called an image
negative, it was correct 81\% of the time (negative predictive value).

\begin{table}[t]
\centering
\caption{Test-Set Confusion Matrix for Image-Level Detection}
\label{tab:confusion}
\small\setlength{\tabcolsep}{4pt}
\begin{tabular}{lcc}
\toprule
 & Predicted negative & Predicted positive \\
\midrule
True negative (1{,}845) & 1{,}672 & 173 \\
True positive (823)     & 392     & 431 \\
\bottomrule
\end{tabular}
\end{table}

\subsection{Qualitative Examples}

\begin{figure}[t]
\centering
\includegraphics[width=0.75\columnwidth]{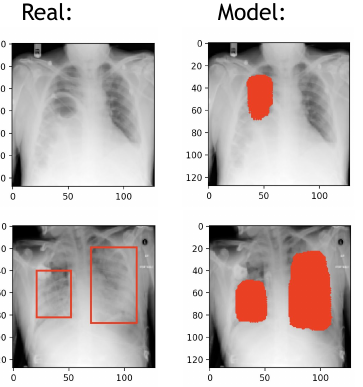}
\caption{Ground truth (left) and predicted masks in red (right) for two
test images. Top: a false positive on an image without pneumonia. Bottom:
predictions that closely match the two radiologist bounding boxes.}
\label{fig:preds}
\end{figure}

Fig.~\ref{fig:preds} shows two test images with the radiologist boxes and
the model's predicted masks. In the first image, which contains no
pneumonia, the model predicted a region in the upper left of the chest, a
false positive. In the second, the model predicted two regions that closely
match the two radiologist-drawn boxes. Because the model was trained on
rectangular masks, its predictions tend to be compact, rounded blobs
rather than anatomically detailed outlines.

\section{Discussion}

\subsection{Does Learning the Harder Task Help?}

The results support the first part of the research question. A model
trained only to segment pneumonia, and never trained to make yes-or-no
decisions, became a usable detector: it identified pneumonia at the image
level with an F1 score of 0.60 and a specificity of 0.91, and its scores
were stable between the development and test sets. It achieves this while
also producing a map of the suspected disease (Fig.~\ref{fig:preds}),
which a classification-only model does not provide.

The higher F1 score on detection (0.60) than on segmentation (0.40--0.45)
is consistent with the idea that a model trained on a harder task finds the
simpler one easier. However, this gap alone does not establish that the
segmentation-first approach is \emph{better} than training for detection
directly. Detection is inherently the easier task for any model: a
prediction that overlaps a radiologist's box only partly still counts as a
correct detection, whereas pixel-level scoring penalizes every boundary
error. The bounding-box masks also lower pixel-level scores, because boxes
include healthy tissue around each opacity. A detection score above the
segmentation score would therefore be expected even without any benefit
from learning segmentation.

The decisive test of the hypothesis is to train a classification model
for detection alone, using the same data splits, image size, and training
budget, and compare its image-level F1 score with the 0.60 achieved here.
A fair version of this experiment would reuse the U-Net's contracting path
as the classifier's backbone, so that the only difference between the two
models is whether they were trained with pixel-level or image-level labels.
That comparison was not completed in this study, so the advantage of
segmentation-first training remains a hypothesis supported by indirect
evidence rather than a demonstrated result.

\subsection{Low Sensitivity}

The model's main weakness is that it missed nearly half of the positive
images. For a screening aid, missed cases are usually more costly than
false alarms, so a clinically useful version would need to trade some
specificity for higher sensitivity. Lowering the detection threshold is
one way to do so. Likely causes of the missed cases include the low
$128\times128$ input resolution, which can erase faint or small opacities;
training from scratch rather than starting from a network pretrained on
images; and the pixel-level imbalance, which persists even with focal
loss. Opacities overlapping the heart, the motivating difficulty for this
work, were not evaluated separately, so it is not known how the model
performs on them specifically.

\subsection{Context}

The F1 scores reported here are not directly comparable with the RSNA
challenge leaderboard, which used mean average precision over box IoU
thresholds and a test set annotated by consensus of multiple radiologists
\cite{gabruseva2020}. They are also not comparable with CheXNet's F1 of
0.435 \cite{rajpurkar2017}, which was measured on a different dataset with
different labels. These results do, however, show that inter-observer
variability limits all methods on this task: even expert radiologists
disagree substantially on which radiographs show pneumonia
\cite{neuman2012,rajpurkar2017}.

\subsection{Limitations}

The model was trained on downsampled images and approximate rectangular
masks, and the development process was limited by training time, so
several configurations were trained for only 20 to 50 epochs. Results come
from a single training run per configuration, without repeated runs to
measure variability. The RSNA labels identify opacities suspicious for
pneumonia as judged by radiologists, not confirmed clinical diagnoses.

\section{Future Work}

\begin{itemize}
\item \emph{Higher resolution.} Training at $256\times256$ or $512\times512$
pixels should preserve faint opacities that are lost at $128\times128$.
\item \emph{Pretrained encoders.} Replacing the U-Net's contracting path
with a network pretrained on large image datasets usually improves
performance when labeled medical data are limited.
\item \emph{Direct test of the detection-only model.} Training a classifier for
detection alone, with the same backbone, data splits, and training budget,
and comparing it with the segmentation-derived detector would directly
answer whether learning the harder task produces a better detector. Also
testing a model trained on both tasks at once would show whether the two
kinds of supervision complement each other.
\item \emph{Loss functions and thresholds.} Combining focal loss with an
overlap-based loss such as Dice loss, and tuning the detection threshold
for higher sensitivity, may reduce missed cases.
\item \emph{Longer training and more data.} Training the selected
configuration for more epochs, with data augmentation and additional
datasets, should reduce overfitting.
\item \emph{Targeted evaluation.} Evaluating separately on opacities behind
the heart and in early disease would show whether the model helps in the
cases radiologists find hardest.
\end{itemize}

\section{Conclusion}

This paper investigated whether a model trained on the harder task of
segmenting pneumonia on chest radiographs can be used for the simpler task
of detecting it. A U-Net trained on the bounding-box annotations of the
RSNA Pneumonia Detection Challenge, using focal loss and dropout to cope
with severe pixel-level class imbalance, reached a pixel-level F1 score of
0.45 on segmentation. Used as a detector, the same model reached an F1
score of 0.60 with 91\% specificity, while also showing where the suspected
disease is located. These results show that segmentation-first training
produces a working detector. Whether it produces a better detector than
training for detection alone is the open question this work sets up, and a
direct comparison with a classification model trained on the same data is
the next step. Improving sensitivity, which was 0.52, through higher input
resolution and pretrained encoders is the other priority on the way to a
tool that could help radiologists identify pneumonia faster and more
reliably.


\end{document}